\documentclass[conference]{IEEEtran}
\IEEEoverridecommandlockouts

\RequirePackage[utf8]{inputenc}
\RequirePackage[english]{babel}
\RequirePackage{cite}
\RequirePackage{amsmath,amssymb,amsfonts}
\RequirePackage{algorithmic}
\RequirePackage[pscoord]{eso-pic}
\RequirePackage{textcomp}
\RequirePackage{xcolor}
\RequirePackage[a4paper, total={184mm,239mm}]{geometry}
\PassOptionsToPackage{detect-all, per-mode=symbol, range-units=single, range-phrase=~to~}{siunitx}
\RequirePackage{siunitx}
\RequirePackage{nicefrac}
\RequirePackage[hidelinks]{hyperref}
\RequirePackage{orcidlink}
\RequirePackage[noabbrev,capitalise]{cleveref}
\RequirePackage{float}
\RequirePackage[caption=false]{subfig}
\RequirePackage{url}
\RequirePackage{lipsum}
\RequirePackage{placeins}
\RequirePackage{xfrac}

\RequirePackage{amssymb}
\RequirePackage{pifont}

\RequirePackage{threeparttable}
\RequirePackage{booktabs}
\RequirePackage{colortbl}
\RequirePackage{tabularx}
\RequirePackage{makecell}

\RequirePackage{graphicx}
\RequirePackage{glossaries}
\RequirePackage{xspace}

\definecolor{ieee-bright-dblue-100}{rgb}{0.0, 0.3828, 0.6055}
\definecolor{ieee-bright-dblue-80}{rgb}{0.0, 0.4883, 0.6797}
\definecolor{ieee-bright-dblue-60}{rgb}{0.3633, 0.6094, 0.7617}
\definecolor{ieee-bright-dblue-40}{rgb}{0.5898, 0.7383, 0.8398}
\definecolor{ieee-bright-dblue-20}{rgb}{0.8906, 0.8984, 0.9219}
\definecolor{ieee-bright-red-100}{rgb}{0.7266, 0.0469, 0.1836}
\definecolor{ieee-bright-red-80}{rgb}{0.832, 0.3164, 0.3281}
\definecolor{ieee-bright-red-60}{rgb}{0.8906, 0.4922, 0.4805}
\definecolor{ieee-bright-red-40}{rgb}{0.9336, 0.6562, 0.6406}
\definecolor{ieee-bright-red-20}{rgb}{0.9688, 0.8203, 0.8125}
\definecolor{ieee-bright-orange-100}{rgb}{0.9961, 0.6367, 0.0}
\definecolor{ieee-bright-orange-80}{rgb}{0.9844, 0.6953, 0.3125}
\definecolor{ieee-bright-orange-60}{rgb}{0.9883, 0.7695, 0.4844}
\definecolor{ieee-bright-orange-40}{rgb}{0.9922, 0.8359, 0.6562}
\definecolor{ieee-bright-orange-20}{rgb}{0.9961, 0.9219, 0.8164}
\definecolor{ieee-bright-yellow-100}{rgb}{0.9961, 0.8164, 0.0}
\definecolor{ieee-bright-yellow-80}{rgb}{0.9961, 0.8477, 0.2148}
\definecolor{ieee-bright-yellow-60}{rgb}{0.9961, 0.875, 0.4492}
\definecolor{ieee-bright-yellow-40}{rgb}{0.9961, 0.9062, 0.6328}
\definecolor{ieee-bright-yellow-20}{rgb}{0.9961, 0.9531, 0.8125}
\definecolor{ieee-bright-lgreen-100}{rgb}{0.4688, 0.7422, 0.125}
\definecolor{ieee-bright-lgreen-80}{rgb}{0.5742, 0.7852, 0.332}
\definecolor{ieee-bright-lgreen-60}{rgb}{0.6875, 0.8398, 0.5039}
\definecolor{ieee-bright-lgreen-40}{rgb}{0.793, 0.8906, 0.6641}
\definecolor{ieee-bright-lgreen-20}{rgb}{0.8945, 0.9414, 0.8281}
\definecolor{ieee-bright-dgreen-100}{rgb}{0.0, 0.5156, 0.2383}
\definecolor{ieee-bright-dgreen-80}{rgb}{0.1641, 0.6055, 0.3867}
\definecolor{ieee-bright-dgreen-60}{rgb}{0.3906, 0.6953, 0.5234}
\definecolor{ieee-bright-dgreen-40}{rgb}{0.6094, 0.8008, 0.6719}
\definecolor{ieee-bright-dgreen-20}{rgb}{0.8047, 0.8945, 0.8359}
\definecolor{ieee-bright-purple-100}{rgb}{0.5938, 0.1133, 0.5898}
\definecolor{ieee-bright-purple-80}{rgb}{0.6992, 0.3281, 0.668}
\definecolor{ieee-bright-purple-60}{rgb}{0.7812, 0.4961, 0.7461}
\definecolor{ieee-bright-purple-40}{rgb}{0.8555, 0.6602, 0.8281}
\definecolor{ieee-bright-purple-20}{rgb}{0.9219, 0.8281, 0.9023}
\definecolor{ieee-bright-lblue-100}{rgb}{0.0, 0.6094, 0.6484}
\definecolor{ieee-bright-lblue-80}{rgb}{0.0, 0.6797, 0.7188}
\definecolor{ieee-bright-lblue-60}{rgb}{0.2109, 0.75, 0.7812}
\definecolor{ieee-bright-lblue-40}{rgb}{0.5469, 0.8242, 0.8438}
\definecolor{ieee-bright-lblue-20}{rgb}{0.7695, 0.918, 0.9219}
\definecolor{ieee-bright-cyan-100}{rgb}{0.0, 0.707, 0.8828}
\definecolor{ieee-bright-cyan-80}{rgb}{0.0, 0.7227, 0.9453}
\definecolor{ieee-bright-cyan-60}{rgb}{0.2656, 0.7812, 0.957}
\definecolor{ieee-bright-cyan-40}{rgb}{0.5547, 0.8438, 0.9688}
\definecolor{ieee-bright-cyan-20}{rgb}{0.7773, 0.9141, 0.9805}
\definecolor{ieee-bright-white-100}{rgb}{0.9961, 0.9961, 0.9961}
\definecolor{ieee-bright-white-80}{rgb}{0.9961, 0.9961, 0.9961}
\definecolor{ieee-bright-white-60}{rgb}{0.9961, 0.9961, 0.9961}
\definecolor{ieee-bright-white-40}{rgb}{0.9961, 0.9961, 0.9961}
\definecolor{ieee-bright-white-20}{rgb}{0.9961, 0.9961, 0.9961}
\definecolor{ieee-dark-red-100}{rgb}{0.5234, 0.1211, 0.2539}
\definecolor{ieee-dark-red-80}{rgb}{0.6445, 0.2812, 0.3828}
\definecolor{ieee-dark-red-60}{rgb}{0.7422, 0.4727, 0.5234}
\definecolor{ieee-dark-red-40}{rgb}{0.832, 0.6445, 0.6758}
\definecolor{ieee-dark-red-20}{rgb}{0.918, 0.8203, 0.832}
\definecolor{ieee-dark-orange-100}{rgb}{0.9062, 0.4648, 0.1328}
\definecolor{ieee-dark-orange-80}{rgb}{0.9648, 0.5664, 0.3164}
\definecolor{ieee-dark-orange-60}{rgb}{0.9766, 0.6758, 0.4805}
\definecolor{ieee-dark-orange-40}{rgb}{0.9844, 0.7773, 0.6523}
\definecolor{ieee-dark-orange-20}{rgb}{0.9922, 0.8789, 0.8125}
\definecolor{ieee-dark-yellow-100}{rgb}{0.9961, 0.7773, 0.1719}
\definecolor{ieee-dark-yellow-80}{rgb}{0.9961, 0.8086, 0.375}
\definecolor{ieee-dark-yellow-60}{rgb}{0.9961, 0.875, 0.4492}
\definecolor{ieee-dark-yellow-40}{rgb}{0.9961, 0.8984, 0.6875}
\definecolor{ieee-dark-yellow-20}{rgb}{0.9961, 0.9453, 0.8438}
\definecolor{ieee-dark-lgreen-100}{rgb}{0.3945, 0.5508, 0.0938}
\definecolor{ieee-dark-lgreen-80}{rgb}{0.5078, 0.6289, 0.293}
\definecolor{ieee-dark-lgreen-60}{rgb}{0.6367, 0.7188, 0.4688}
\definecolor{ieee-dark-lgreen-40}{rgb}{0.7539, 0.8047, 0.6367}
\definecolor{ieee-dark-lgreen-20}{rgb}{0.875, 0.9023, 0.8125}
\definecolor{ieee-dark-dgreen-100}{rgb}{0.0, 0.3867, 0.2539}
\definecolor{ieee-dark-dgreen-80}{rgb}{0.1836, 0.5, 0.3906}
\definecolor{ieee-dark-dgreen-60}{rgb}{0.3984, 0.6172, 0.5273}
\definecolor{ieee-dark-dgreen-40}{rgb}{0.5938, 0.7422, 0.6758}
\definecolor{ieee-dark-dgreen-20}{rgb}{0.793, 0.8711, 0.8359}
\definecolor{ieee-dark-purple-100}{rgb}{0.4648, 0.1445, 0.5117}
\definecolor{ieee-dark-purple-80}{rgb}{0.5898, 0.3242, 0.6016}
\definecolor{ieee-dark-purple-60}{rgb}{0.6914, 0.4883, 0.6953}
\definecolor{ieee-dark-purple-40}{rgb}{0.7969, 0.6523, 0.793}
\definecolor{ieee-dark-purple-20}{rgb}{0.8945, 0.8203, 0.8945}
\definecolor{ieee-dark-cyan-100}{rgb}{0.0, 0.4492, 0.4648}
\definecolor{ieee-dark-cyan-80}{rgb}{0.0, 0.5469, 0.5664}
\definecolor{ieee-dark-cyan-60}{rgb}{0.3047, 0.6602, 0.668}
\definecolor{ieee-dark-cyan-40}{rgb}{0.5586, 0.7695, 0.7734}
\definecolor{ieee-dark-cyan-20}{rgb}{0.7734, 0.8789, 0.8789}
\definecolor{ieee-dark-dblue-100}{rgb}{0.0, 0.1562, 0.332}
\definecolor{ieee-dark-dblue-80}{rgb}{0.1797, 0.3008, 0.4609}
\definecolor{ieee-dark-dblue-60}{rgb}{0.3828, 0.4609, 0.5859}
\definecolor{ieee-dark-dblue-40}{rgb}{0.5781, 0.6289, 0.7188}
\definecolor{ieee-dark-dblue-20}{rgb}{0.7852, 0.8047, 0.8555}
\definecolor{ieee-dark-grey-100}{rgb}{0.457, 0.4688, 0.4805}
\definecolor{ieee-dark-grey-80}{rgb}{0.5625, 0.5625, 0.5742}
\definecolor{ieee-dark-grey-60}{rgb}{0.6641, 0.6641, 0.6758}
\definecolor{ieee-dark-grey-40}{rgb}{0.7734, 0.7695, 0.7773}
\definecolor{ieee-dark-grey-20}{rgb}{0.8789, 0.8828, 0.8828}
\definecolor{ieee-dark-black-100}{rgb}{0.0, 0.0, 0.0}
\definecolor{ieee-dark-black-80}{rgb}{0.3438, 0.3477, 0.3555}
\definecolor{ieee-dark-black-60}{rgb}{0.5, 0.5078, 0.5195}
\definecolor{ieee-dark-black-40}{rgb}{0.6523, 0.6602, 0.6719}
\definecolor{ieee-dark-black-20}{rgb}{0.8164, 0.8242, 0.8281}

\definecolor{light-gray}{gray}{0.75}

\newcommand*\circnum[1]{\tikz[baseline=(char.base)]{%
            \node[white,shape=circle,fill=ieee-dark-black-100,draw,inner sep=1pt] (char) {\color{ieee-bright-white-100}\sffamily #1};}}

\newcommand{\glsu}[1]{\glsunset{#1}\gls{#1}}

\newcommand{\etal}{\emph{et al.}}
\newcommand{\x}{$\times$}

\renewcommand{\subsubsection}[1]{\paragraph*{\textbf{#1}}}

\DeclareSIUnit{\x}{\!\ensuremath{\times}}
\DeclareSIUnit\bit{b}
\DeclareSIUnit\GE{GE}
\DeclareSIUnit\kGE{\kilo\GE}
\DeclareSIUnit\MGE{\mega\GE}
\def\reviewpass{v4.0.0}
\def\axirt{AXI-REALM\xspace} %

\begin{document}

\AddToShipoutPictureBG*{%
  \AtPageUpperLeft{%
    \hspace{\paperwidth}%
    \raisebox{-\baselineskip}{%
      \makebox[-35pt][r]{\footnotesize{
        \copyright~2023~IEEE. Personal use of this material is permitted. %
        Permission from IEEE must be obtained for all other uses, in any current or future media, including
      }}
}}}%

\AddToShipoutPictureBG*{%
  \AtPageUpperLeft{%
    \hspace{\paperwidth}%
    \raisebox{-2\baselineskip}{%
      \makebox[-37pt][r]{\footnotesize{
        reprinting/republishing this material for advertising or promotional purposes, creating new collective works, for resale or redistribution to servers or lists, or
      }}
}}}%

\AddToShipoutPictureBG*{%
  \AtPageUpperLeft{%
    \hspace{\paperwidth}%
    \raisebox{-3\baselineskip}{%
      \makebox[-185pt][r]{\footnotesize{
       reuse of any copyrighted component of this work in other works.
      }}
}}}%

\ifx\showrevision\undefined
    \newcommand{\todo}[1]{{#1}}
    \newcommand{\ph}[1]{{#1}}
\else
    \newcommand{\todo}[1]{{\textcolor{ieee-bright-red-80}{#1}}\PackageWarning{TODO:}{#1!}}
    \newcommand{\ph}[1]{{\textcolor{light-gray}{#1}}\PackageWarning{PH:}{#1!}}
    \AddToShipoutPictureFG{%
        \put(%
            8mm,%
            \paperheight-1.5cm%
            ){\vtop{{\null}\makebox[0pt][c]{%
                \rotatebox[origin=c]{90}{%
                    \huge\textcolor{ieee-bright-red-80!75}{\reviewpass}%
                }%
            }}%
        }%
    }
    \AddToShipoutPictureFG{%
        \put(%
            \paperwidth-6mm,%
            \paperheight-1.5cm%
            ){\vtop{{\null}\makebox[0pt][c]{%
                \rotatebox[origin=c]{90}{%
                    \huge\textcolor{ieee-bright-red-80!30}{ETH Zurich - Unpublished - Confidential - Draft - Copyright 2023}%
                }%
            }}%
        }%
    }
\fi

\newacronym{sota}{SOTA}{state-of-the-art}
\newacronym{fpga}{FPGA}{field programmable gate array}
\newacronym{asic}{ASIC}{application-specific integrated circuit}
\newacronym{fub}{FUB}{functional unit block}
\newacronym{vv}{V\&V}{validation and verification}
\newacronym{gpp}{GPP}{general purpose processor}

\newacronym{hpc}{HPC}{high performance computing}
\newacronym{ml}{ML}{machine learning}
\newacronym{isa}{ISA}{instruction set architecture}
\newacronym{fp}{FP}{floating-point}
\newacronym{dl}{DL}{deep learning}
\newacronym{la}{LA}{linear algebra}
\newacronym{ip}{IP}{intellectual property}
\newacronym[firstplural=systems-on-chip (SoCs)]{soc}{SoC}{system-on-chip}
\newacronym{mpsoc}{MPSoC}{multi-processor system-on-chip}
\newacronym[firstplural=networks-on-chip (NoCs)]{noc}{NoC}{network-on-chip}
\newacronym{hw}{HW}{hardware}
\newacronym{sw}{SW}{software}
\newacronym{swapc}{SWaP-C}{space, weight, power, and cost}
\newacronym{mcp}{MCP}{multi-core processor}
\newacronym{rr}{RR}{round-robin}

\newacronym{mac}{MAC}{multiply-accumulate}
\newacronym{fem}{FEM}{finite element analysis}
\newacronym{simd}{SIMD}{single-instruction, multiple-data}
\newacronym{rtl}{RTL}{register transfer level}
\newacronym{dlt}{DLT}{data layout transform}

\newacronym{fifo}{FIFO}{first in, first out}
\newacronym{fu}{FU}{functional unit}
\newacronym{alu}{ALU}{arithmetic logic unit}
\newacronym{fpu}{FPU}{floating-point unit}
\newacronym{ssr}{SSR}{stream semantic register}
\newacronym{issr}{ISSR}{indirection stream semantic register}
\newacronym{tcdm}{TCDM}{tightly-coupled data memory}
\newacronym{dma}{DMA}{direct memory access}
\newacronym{sm}{SM}{streaming multiprocessor}
\newacronym{vlsu}{VLSU}{vector load-store unit}
\newacronym{dsa}{DSA}{domain-specific accelerator}
\newacronym{ha}{HA}{hardware accelerator}
\newacronym{fsm}{FSM}{finite state machine}
\newacronym{llc}{LLC}{last-level cache}
\newacronym{d2d}{D2D}{die-to-die}
\newacronym{dram}{DRAM}{dynamic random access memory}
\newacronym{tid}{TID}{transaction ID}
\newacronym{spm}{SPM}{scratchpad memory}

\newacronym{os}{OS}{operating system}

\newacronym{spvv}{SpVV}{sparse vector-vector multiplication}
\newacronym{spmv}{SpMV}{sparse vector-matrix multiplication}
\newacronym{spmm}{SpMM}{sparse matrix-matrix multiplication}
\newacronym{csrmv}{CsrMV}{CSR matrix-vector multiplication}
\newacronym{csrmm}{CsrMM}{CSR matrix-matrix multiplication}

\newacronym{csf}{CSF}{compressed sparse fiber}
\newacronym{csr}{CSR}{compressed sparse rows}
\newacronym{csc}{CSC}{compressed sparse columns}
\newacronym{bcsr}{BCSR}{blocked compressed sparse rows}

\newacronym{axi4}{AXI4}{Advanced eXtensible Interface 4}
\newacronym{amba}{AMBA}{Advanced Microcontroller Bus Architecture}
\newacronym{sram}{SRAM}{static random-access memory}

\newacronym{wcet}{WCET}{worst-case execution time}
\newacronym{rtunit}{REALM unit}{real-time regulation and traffic monitoring unit}
\newacronym{mtunit}{M\&R unit}{monitoring and regulation unit}
\newacronym{cps}{CPS}{cyber-physical system}
\newacronym{crtes}{CRTES}{critical real-time embedded system}
\newacronym{heicps}{He-iCPS}{heterogeneous integrated cyber-physical system}
\newacronym{ecu}{ECU}{electronic control unit}
\newacronym{mcs}{MCS}{mixed criticality system}
\newacronym{ima}{IMA}{integrated modular avionics}
\newacronym{adas}{ADAS}{advanced driver assistance system}
\newacronym{axirealm}{AXI-REALM}{AXI real-time regulation and traffic monitoring}
\newacronym{mpam}{MPAM}{memory system resource partitioning and monitoring}
\newacronym{dos}{DoS}{denial of service}
\newacronym{hwrot}{HWRoT}{hardware root of trust}

\title{{\axirt}: A Lightweight and Modular \\ Interconnect Extension for Traffic Regulation and Monitoring of Heterogeneous Real-Time SoCs}

\ifx\blind\undefined
    \author{
        \IEEEauthorblockN{%
        Thomas Benz\orcidlink{0000-0002-0326-9676}\IEEEauthorrefmark{1}\IEEEauthorrefmark{10}, %
        Alessandro Ottaviano\orcidlink{0009-0000-9924-3536}\IEEEauthorrefmark{1}\IEEEauthorrefmark{10}, %
        Robert Balas\orcidlink{0000-0002-7231-9315}\IEEEauthorrefmark{1}, %
        Angelo Garofalo\orcidlink{0000-0000-0000-0000}\IEEEauthorrefmark{1}\IEEEauthorrefmark{2}, \\%
        Francesco Restuccia\orcidlink{0000-0000-0000-0000}\IEEEauthorrefmark{3}, %
        Alessandro Biondi\orcidlink{0000-0000-0000-0000}\IEEEauthorrefmark{4}, %
        Luca Benini\orcidlink{0000-0001-8068-3806}\IEEEauthorrefmark{1}\IEEEauthorrefmark{2}%
        }
        \thanks{%
            \IEEEauthorrefmark{10} Both authors contributed equally to this research.
        }
        \IEEEauthorblockA{
            \IEEEauthorrefmark{1}~\textit{Integrated Systems Laboratory, ETH Zurich}, Switzerland \\
            \IEEEauthorrefmark{2}~\textit{Department of Electrical, Electronic, and Information Engineering, University of Bologna}, Italy \\
            \IEEEauthorrefmark{3}~\textit{Department of Computer Science and Engineering, UC San Diego}, San Diego, CA USA \\
            \IEEEauthorrefmark{4}~\textit{Department of Excellence in Robotics \& AI, Scuola Superiore Sant'Anna}, Pisa, Italy \\
        }
    }
\else
    \author{%
            \vspace{1.1cm} %
            \textit{Authors omitted for blind review}
            \vspace{1.1cm} %
            }
\fi

\maketitle

\begin{abstract}
The increasing demand for heterogeneous functionality in the automotive industry and the evolution of chip manufacturing processes have led to the transition from federated to integrated critical real-time embedded systems (CRTESs). %
This leads to higher integration challenges of conventional timing predictability techniques due to access contention on shared resources, which can be resolved by providing system-level observability and controllability in hardware. %
We focus on the interconnect as a shared resource and propose {\axirt}, a lightweight, modular, and technology-independent real-time extension to industry-standard AXI4 interconnects, available open-source. %
{\axirt} uses a credit-based mechanism to distribute and control the bandwidth in a multi-subordinate system on periodic time windows, proactively prevents denial of service from malicious actors in the system, and tracks each manager's access and interference statistics for optimal budget and period selection. %
We provide detailed performance and implementation cost assessment in a 12nm node and an end-to-end functional case study implementing {\axirt} into an open-source Linux-capable RISC-V SoC. %
In a system with a general-purpose core and a hardware accelerator's DMA engine causing interference on the interconnect, {\axirt} achieves fair bandwidth distribution among managers, allowing the core to recover \SI{68.2}{\%} of its performance compared to the case without contention.
Moreover, near-ideal performance (above \SI{95}{\%}) can be achieved by distributing the available bandwidth in favor of the core, improving the worst-case memory access latency from 264 to below eight cycles.
Our approach minimizes buffering compared to other solutions and introduces only \SI{2.45}{\%} area overhead compared to the original SoC.
\end{abstract}


\section{Introduction}
\label{sec:intro}

The growing demand for heterogeneous functionalities with mixed criticality features~\cite{RETIS_BANDWIDTH_RESERVATION, AXIICRT_ARM} has led to a \emph{new renaissance} in the design of \glspl{crtes} such as modern cars in recent years.
\emph{Trusted} (safety- and time-critical) tasks typically include engine, brake, and cruise control, \gls{adas}, and telematics~\cite{MCKINSEY_AUTOMOTIVE_SURVEY}. 
These systems comprise single or multi-core general-purpose processors requiring strict certification requirements enforced through timing predictability and composability analysis~\cite{SOA_WCET}.
\emph{Untrusted} (soft time-critical) computational tasks run infotainment and commodity applications~\cite{MCKINSEY_AUTOMOTIVE_SURVEY}. 
Untrusted systems are either processors running computational tasks or independent \glspl{dsa} running memory-intensive workloads, such as \gls{ml} for \gls{adas}~\cite{DEEP_LEARNING_ADAS}.
Their applications typically require softer timing bounds and are often not subject to certification processes. 

The demand for such heterogeneous functionalities has dramatically raised the complexity of modern real-time \glspl{cps}~\cite{EVIDENCE_ETH_TOWARDS_OPEN_AUTOMOTIVE_PLATFORM}.
The number of \glspl{ecu} in cars has grown to as much as 150 per vehicle, but the continuous addition of \gls{hw} has become untenable due to \gls{swapc} constraints, especially in terms of vehicle cable harnesses~\cite{MCKINSEY_AUTOMOTIVE_SURVEY}. %

At the same time, the evolution of manufacturing processes has made available more performant \glspl{soc} to satisfy such diverse functionality requirements.
For this reason, novel real-time architectures in the automotive domain are moving towards merging federated \glspl{ecu} on the same \gls{hw} platform~\cite{RETIS_SPATIAL_TEMPORAL_PARTITIONING}. 
Similarly to \gls{ima} architectures in airborne systems~\cite{IMA_AIRBORNE}, the platform becomes an integrated, heterogeneous \gls{mcs}. %

While this solution mitigates the \gls{swapc} problem, it makes timing predictability and composability analysis more challenging because of the increased interference generated by multiple actors (critical and non-critical \glspl{mcp} clusters, \glspl{ha}, \gls{dma} engines) coexisting on the same platform and contending for shared \gls{hw} resources.
If not considered, the additional contention may introduce unpredictable behavior during the system's execution, causing possible deadline misses for trusted tasks~\cite{RETIS_AXI_HYPER_CONNECT, CAST32A_POSITION_PAPER}.
To preserve the timing behavior of the system under known and predictable bounds, enhancing \emph{observability} and \emph{controllability} of the shared \gls{hw} resources through temporal isolation becomes a prerequisite~\cite{AUTOMOTIVE_PREDICTABLE}.
Of particular interest is the interconnect of modern \glspl{soc}. 

Some existing \gls{hw} solutions demand intrusive changes to the interconnect, requiring further iterations on the verification process of the \gls{ip}~\cite{RETIS_AXI_HYPER_CONNECT, AXIICRT_ARM}.
Other approaches are less invasive but lack a cohesive design that addresses both \emph{observability} and \emph{controllability} of the interconnect~\cite{RETIS_BANDWIDTH_RESERVATION, RETIS_IS_BUS_ARBITER_FAIR}.
Furthermore, to our best knowledge, all \gls{sota} design proposals limit their cost and functionality assessments to \glspl{fpga} (\Cref{sec:relwork}).

\subsubsection{Contribution}
We address the mentioned limitations and propose {\axirt}, a modular, lightweight, and implementation-agnostic real-time extension to system bus supporting \gls{axi4}, the \emph{de facto} standard for on-chip, non-coherent \gls{soc} interconnects.
The synthesizable \glsu{rtl} description of {\axirt} is available free and %
\ifx\blind\undefined
     open-source~\footnote{\texttt{\url{https://github.com/pulp-platform/axi_rt}}}.
\else
     open-source~\footnote{\texttt{URL omitted for blind review}}.
\fi
We present the following contributions: 
\begin{enumerate}

    \item \textbf{\Gls{hw}-driven Traffic Controllability.} {\axirt} implements a per-manager configurable number of subordinate \emph{regions}, with runtime-configurable address range, budget, and reservation period per \emph{region}. 
    Each manager can be isolated from the system in case of malicious attacks, and a \textit{bus guard} module (\Cref{sec:arch}) allows privileged access to the unit's configuration registers.

    \item \textbf{\Gls{hw}-driven Traffic Observability.} A \gls{mtunit} tracks per-manager access and interference statistics, such as transaction latency, bandwidth, and interference with each other manager, and eventually performs per-manager budget-aware bandwidth throttling within the period, see \Cref{sec:arch:mtu}.
    
    \item \textbf{Implementation-agnostic Design.} We integrate {\axirt} in Cheshire, a Linux-capable RISC-V \gls{soc}~\cite{10163410}, and characterize it in a \SI{12}{\nano\metre} technology. 
    {\axirt} incurs an area overhead of only \SI{2.45}{\%} at isofrequency compared to the original design. 
    We can improve a memory-intense benchmark from \SI{0.7}{\%} to \SI{68.2}{\%} of the \emph{single-source} performance with equal budget distribution and \emph{near-ideal} performance when distributing the budget in favor of the core, improving the worst-case memory access latency from over \emph{264} to below \emph{eight} cycles.
    
\end{enumerate}

\section{Related Work}
\label{sec:relwork}

\Gls{sota} techniques to improve the real-time capabilities of heterogeneous \gls{soc} interconnects traditionally follow two design strategies: adding real-time regulation modules between the managers and the interconnect itself or intrusive customization of existing interconnect architectures.
Furthermore, \gls{sota} solutions restrict the design and evaluation of \glspl{fpga} or \gls{fpga}-\glspl{soc}, lacking an open-source hardware platform enabling gate-level characterization in a modern technology node.

\subsubsection{Regulation Helper Modules}
\label{subsec:relwork_helpers}

Several works propose credit-based mechanisms to enforce spatial and temporal upper bounds to non-coherent, on-chip interconnect networks.

Pagani~{\etal} and Restuccia~{\etal} analyze and address the problem of multiple \glspl{dsa} competing for bandwidth or causing \emph{\gls{dos}} in heterogeneous, \gls{axi4}-based \gls{fpga}-\glspl{soc} and propose three units to mitigate contention.
The AXI budgeting unit (ABU)~\cite{RETIS_BANDWIDTH_RESERVATION} extends the concept of inter-core memory reservation on \glspl{mpsoc} to heterogeneous \glspl{soc} and proposes counter-based budgets and periods assigned to each manager in the system.
The AXI burst equalizer (ABE)~\cite{RETIS_IS_BUS_ARBITER_FAIR} tackles unfair arbitration and bandwidth stealing from malicious managers by enforcing a nominal burst size and maximum number of outstanding transactions per manager basis.
Finally, the Cut and Forward (C\&F)~\cite{RETIS_CUT_FORWARD} unit moves the burden of completing a transaction from an untrusted manager to the interconnect. %
A misbehaving manager could cause \gls{dos} by reserving write bandwidth and never completing the transaction (stalling). 
{\axirt} tackles these challenges by optimizing the design for high performance, adding only one cycle of latency (~\Cref{sec:arch}) and extremely low area overhead (\Cref{sec:eval:sys}).

A crucial feature of temporal isolation on shared resources is extracting meaningful information from the functional units during validation, allowing the selection of effective maximum bounds during operation.
In~\cite{SafeSU2021}, Cabo~{\etal} propose SafeSU, a minimally invasive statistics unit. 
SafeSU is limited to tracking inter-core interference in \glspl{mpsoc}. 
We extend monitoring capabilities in {\axirt} to heterogeneous \glspl{soc}, tracking average bandwidth, latency, and inter-\gls{dsa} interference.

\subsubsection{Interconnect Customization}
\label{subsec:relwork_bus_redesign}

This strategy profoundly changes the \glspl{ip}'s inner dataflow.
Restuccia~{\etal}~\cite{RETIS_AXI_HYPER_CONNECT} propose HyperConnect, a custom \gls{axi4}-based \gls{fub} for virtualized \gls{fpga}-\glspl{soc}. While being the closest \gls{sota} to {\axirt}, HyperConnect cannot prevent \gls{dos} from a misbehaving manager stalling the interconnect (\Cref{sec:arch:buf}).

Recently, Jiang~{\etal} introduced AXI-IC$^{RT}$~\cite{AXIICRT_ARM}, one of the first end-to-end \gls{axi4} microarchitectures tailored for real-time use cases. 
AXI-IC$^{RT}$ uses the \gls{axi4} user signal to define priorities and proposes a two-layer scheduler algorithm to efficiently select the budget and period for each manager at runtime. 
Instead, {\axirt} does not introduce the concept of priority, which may lead to request starvation on low-priority managers. It relies on a credit-based mechanism and a \emph{granular burst splitter} to distribute the bandwidth according to the real-time guarantee of the \gls{soc}. 
While AXI-IC$^{RT}$ evaluates budget reservation strategies, it limits the assessment to managers with equal credit (bandwidth).
From an implementation angle, the design strategy followed by AXI-IC$^{RT}$ adds extensive buffering to the microarchitecture to create an observation window for early service of incoming transactions based on priorities.
Overall, we observe that both HyperConnect and AXI-IC$^{RT}$ lack monitoring capabilities to track traffic statistics.

In industry, Arm's CoreLink QoS-400 is integrated into modern \gls{fpga}-\glspl{soc} and controls contention using the QoS signal available in the \gls{axi4} and AXI5 specifications.
QoS-400 has several limitations: in a Zynq Ultrascale + \gls{fpga}, the authors report more than 30 QoS points that must work coordinately to control the traffic~\cite{BSC_ARM_QOS400}. 
In the Arm ecosystem, QoS mechanisms are being replaced by \gls{mpam}, a promising industry-grade specification with several guidelines to implement memory access regulation mechanisms~\cite{RETIS_ARM_MPAM}. \Gls{mpam} priority partitioning could be applied to {\axirt}'s flexible configuration interface to support bandwidth control from hypervisors integrating \gls{mpam} discovery mechanisms. 

\section{Architecture}
\label{sec:arch}

\begin{figure}[t]
    \centering
    \subfloat[]{{\includegraphics[height=2.7cm]{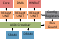}}}%
    \qquad
    \subfloat[]{{\includegraphics[height=2.7cm]{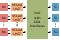}}}%
    \caption{%
        Integration of {\axirt} in a system with a crossbar and a \gls{noc}.
    }
    \label{fig:rt-sys}
\end{figure}

We design {\axirt} to be independent of the memory system's architecture, making it compatible with any memory system featuring \gls{axi4} interfaces, from commonly used crossbar-based interconnects~\cite{arm_amba_2023, 9522037, RETIS_AXI_HYPER_CONNECT} to more scalable \glspl{noc}. %
\Cref{fig:rt-sys} shows two example integrations of the \glspl{rtunit}. 
We focus on crossbar-based interconnects.

Burst-based interconnect architectures, usually \gls{rr}-arbitrate transactions on burst granularity~\cite{9522037}. 
This strategy affects bandwidth distribution fairness by increasing the completion latency of fine-granular transfers in the presence of long bursts issued from \glspl{dsa} in heterogeneous \glspl{soc}. 
Contrarily to safety- and time-critical tasks executed on processors, \glspl{dsa} often work independently~\cite{RETIS_BANDWIDTH_RESERVATION} and can unpredictably cause interference when strict timing bounds are required. %

{ \axirt} regulates and monitors the traffic in three stages: \circnum{1} it employs a period- and budget-based scheme to limit manager accesses with predetermined bandwidth on a given time window, a required feature to achieve timing analysis guarantees in real-time systems (\gls{mtunit}, \Cref{sec:arch:mtu}), \circnum{2} it fragments incoming bursts to a defined granularity, tuned to match the length of the latency-critical transfers (granular burst splitter, \Cref{sec:arch:bs}), and \circnum{3} it forwards write transactions to prevent \gls{dos} from malicious managers (write buffer, \Cref{sec:arch:buf}). %
Compared to existing solutions, budget and period are assigned to a configurable number of subordinate \emph{regions} associated with each manager, providing more flexibility in terms of bandwidth distribution and problem modeling (\Cref{sec:relwork}).
Monitoring, modifying, and controlling each manager's traffic at the ingress into the network enables throttling or stalling a stream with minimal congestion and without additional buffering inside the network.
{\axirt} delays in-flight transactions by just one clock cycle.
This is negligible, as well-designed \gls{axi4} \glspl{ip} are insensitive to latency~\cite{9522037}. %
In the remainder of this section, we provide an in-depth description of {\axirt} microarchitecture.

\subsection{{\axirt} Unit}
\label{sec:arch:rtunit}

\begin{figure}[t]
    \centering
    \includegraphics[width=0.92\columnwidth]{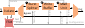}
    \caption{%
        The internal structure of the \gls{rtunit}.
    }
    \label{fig:rt-unit}
\end{figure}

The \gls{rtunit} consists of four sub-blocks orchestrated by a \gls{fsm}, see \Cref{fig:rt-unit}. %
At the ingress of the module, an isolation block allows the manager to be cut off from the memory system, a beneficial feature in case of misbehaving managers~\cite{RETIS_AXI_HYPER_CONNECT}. 
Isolation is triggered under various conditions, including budget depletion, reconfiguration of intrusive parameters, or user-commanded manager isolation. %
The isolation block is aware of any outstanding transactions. %
In the case of user-commanded isolation, it blocks any new transaction while letting outstanding transactions complete. %

Before runtime, an \gls{os} or host hypervisor configures the unit with a period and a budget. 
Every period, the managers receive \emph{transfer budgets} in bytes, which is spent whenever data is transmitted. %
If the budget is depleted, the manager is prevented from further executing any more memory accesses until the period expires and the budget is replenished. %

\subsubsection {Granular Burst Splitter}
\label{sec:arch:bs}

\begin{figure}[t]
    \centering
    \subfloat[\label{fig:rt-splitter}]{{\includegraphics[height=2.8cm]{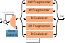}}}
    \qquad
    \subfloat[\label{fig:rt-buffer}]{{\includegraphics[height=3.1cm]{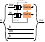}}}
    \caption{%
        The internal architecture of the granular burst splitter (a) and write transaction buffer (b). %
    }
    \label{fig:rt-split-buf}
\end{figure}

This unit fragments incoming burst transactions with runtime-configurable granularity. 
{\axirt} supports burst fragmentation according to the \gls{axi4}~\cite{arm_amba_2023} specification. 
For instance, atomic bursts and \emph{non-modifiable} transactions of length sixteen or smaller cannot be fragmented. %
\Cref{fig:rt-splitter} shows the unit's microarchitecture. %
We store the meta-information of a burst transaction, emit the corresponding fragmented transactions, and update and address information. %
Write responses of the fragmented bursts are coalesced according to the \gls{axi4} specification~\cite{arm_amba_2023}. %
Read responses are passed through, except for the \emph{r.last} signal, which is gated according to the length of the original transaction. %

The splitting granularity is runtime-configurable from one to 256 beats if the write buffer, see \Cref{sec:arch:buf}, is parametrized large enough or is not present. %
If a manager only emits single-word transactions, the granular burst splitter can be disabled from the {\axirt} unit to reduce the area footprint. %

\subsubsection {Write Buffer}
\label{sec:arch:buf}

Most interconnect architectures will reserve the bandwidth for an entire write transaction on the \emph{W} channel once the corresponding \emph{AW} is received~\cite{9522037}. %
A manager device can reserve large amounts of data and stall the interconnect by delaying the corresponding data. 
A malicious manager can \emph{intentionally} enforce this mechanism to cause \gls{dos} of the system, as discussed in~\cite{RETIS_CUT_FORWARD}. 
We prevent this behavior by storing the fragmented write burst in a \emph{write buffer}, \Cref{fig:rt-buffer}. %
The buffer forwards the \emph{AW} request, and the \emph{W} burst once the write data is fully contained within the buffer. %

Read transactions are just passed through, as we assume all of the subordinate devices are returning data in an orderly fashion. %
The transaction buffer is configured to hold two \emph{AWs} and one fragmented write burst, imposing the largest fragmentation size supported by the granular burst splitter. %

\subsubsection {Monitoring and Regulation Unit}
\label{sec:arch:mtu}

\begin{figure}[t]
    \centering
    \includegraphics[width=0.92\columnwidth]{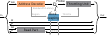}
    \caption{%
        \gls{mtunit}: monitors every region and isolates the corresponding manager if the budget is depleted. %
        Read and write transactions are handled equally; the figure only shows the write part exploded. %
    }
    \label{fig:rt-mon}
\end{figure}

The \gls{mtunit} tracks the transmitted data volume and regulates downstream access with a runtime-configurable period, transfer budget, and address range for each subordinate \emph{region}. 
The regions are defined independently of the physical routing mechanism of the downstream interconnect.
The number of regions is configured at design time. 
\gls{mtunit}'s main benefit is increasing the amount of information the \gls{soc}~\cite{AUTOMOTIVE_PREDICTABLE} exposes to enforce \emph{observability} and \emph{controllability} directly in \gls{hw}. %

\Cref{fig:rt-mon} shows the write path of the unit exploded, whereas the read path is just implied. %
The unit analyzes the address of the transaction, determines the corresponding \emph{region}, and subtracts the number of bytes from the region's budget. %
If at least one of the regions has no budget left, the manager interface is isolated until the budget is replenished, which happens periodically according to the configured reservation period. %
An optional \emph{throttling unit} can be activated, which limits the number of outstanding transactions to the downstream memory system depending on the remaining budget, modulating backpressure before the budget fully expires. %

The \gls{mtunit} provides extensive capabilities for online observability. %
Statistics on the amount of transferred data and elapsed time are measured in relation to the start of the current period, which allows the user to retrieve the region's transfer bandwidth trivially. %
We further provide average latency measurement capabilities within the bookkeeping units. %
An increase in average transaction latency indicates contention in the downstream memory system or the subordinate devices. %
Reading the evolution of the latency from all managers' \glspl{mtunit} and analyzing their statistics provides a full view of the memory system's congestion, extending existing solutions limited to inter-core interference in \glspl{mpsoc} (\Cref{sec:relwork}). %

\subsection{System Integration and Configuration}
\label{sec:cfg}

Depending on the memory system's layout, a shared configuration interface serves one or more \glspl{rtunit}, as shown in \Cref{fig:rt-sys}. %
In the case of a single \gls{axi4} crossbar integrated at the system level, we use a centralized configuration interface to serve all units.  %
To protect against misbehaving or malicious managers, we propose a \emph{bus guard} unit to restrict unwanted access to the configuration interface. %
After a system reset, a trusted manager must claim ownership of the configuration space by writing to a \emph{guard register} within the \emph{bus guard}. In the unclaimed state, every access to the configuration space except for the \emph{guard register} returns an error response. %
Once a manager has claimed an address space within the system, it can perform a \emph{handover} operation to transfer the exclusive read/write ownership to any other manager in the system. The \emph{bus guard} differentiates between managers using their unique \gls{tid}. %
If a \gls{hwrot} is present in the system, we propose that it claims the configuration register during its boot sequence. %
Should a more secure approach be required, the \axirt configuration file can directly and exclusively be connected to the \gls{hwrot}. %

\section{Evaluation}
\label{sec:eval}

\begin{table*}[ht!]
    \centering
    \scriptsize{%
        \centering
        \caption{%
            Area decomposition of the Cheshire \gls{soc}.%
        }
        \label{tab:area-soc}
        \renewcommand*{\arraystretch}{0.95}
        \begin{threeparttable}
            \begin{tabular}{cccccccccccc} \toprule
                \textbf{Unit} & %
                \textbf{SoC} & %
                \textbf{CVA6} & %
                \textbf{LLC} & %
                \textbf{Interconnect~\tnote{a}} & %
                \textbf{\emph{3} RT Units~\tnote{b}} & %
                \textbf{RT CFG} & %
                \textbf{Peripherals} & %
                \textbf{iDMA} & %
                \textbf{Bootrom} & %
                \textbf{IRQ subsys} & %
                \textbf{Rest} \\ %
                
                \midrule
                
                Area {[}kGE{]} &
                3810 & %
                1860 & %
                1350 & %
                206 & %
                \textbf{83.6} & %
                \textbf{9.8} & %
                163 & %
                26.3 & %
                12.9 &  %
                11.1 & %
                20.5 \\ %
                
                Area {[}\%{]} & %
                100.00 & %
                48.7 & %
                35.5 & %
                5.41 & %
                \textbf{2.19} & %
                \textbf{0.26} & %
                4.27 & %
                0.69 & %
                0.34 & %
                0.29 & %
                0.54  \\ %

                \bottomrule
                
            \end{tabular}

            \begin{tablenotes}[para, flushleft]
                \item[a] Without the {\axirt} extension accounted
                \item[b] All 3 units are equally parameterized: \SI{64}{\bit} address and data width, a write buffer depth of \emph{16} elements, \emph{eight} outstanding transfers, and \emph{two} available address regions.
            \end{tablenotes}
        \end{threeparttable}
    }
\end{table*}

\begin{table*}[ht!]
    \centering
    \scriptsize{%
        \centering
        \caption{%
            Area contributions of {\axirt}'s sub-blocks as a function of its parameterization.\\%
            All numbers are in \si{\GE}, at \SI{1}{\giga\hertz} using typical conditions.%
        }%
        \label{tab:ooc-area}
        \renewcommand*{\arraystretch}{0.95}
        \begin{threeparttable}
            \begin{tabular}{l!{\color{black}\vrule}ccccc!{\color{black}\vrule}cccccc} \toprule

                Parameter & 
                \multicolumn{5}{c|}{Configuration Register File} &
                \multicolumn{6}{c}{\gls{rtunit}} \\

                &
                \multicolumn{1}{c!{\color{gray}\vrule}}{Per-system} &
                \multicolumn{1}{c!{\color{gray}\vrule}}{Per-unit} &
                \multicolumn{3}{c!{\color{black}\vrule}}{Per-unit \& Region} &
                \multicolumn{4}{c!{\color{gray}\vrule}}{Per-unit} &
                \multicolumn{2}{c}{{Per-unit \& region}} \\

                &
                \multicolumn{1}{c!{\color{gray}\vrule}}{Bus Guard} &
                \multicolumn{1}{c!{\color{gray}\vrule}}{\begin{tabular}[c]{@{}c@{}}Burst config\\ Register\end{tabular}} &
                \begin{tabular}[c]{@{}c@{}}C\&S\\ Register\end{tabular} &
                \begin{tabular}[c]{@{}c@{}}Budget \& Period\\ Register\end{tabular} &
                \begin{tabular}[c]{@{}c@{}}Region Boundary\\ Register\end{tabular} &
                \begin{tabular}[c]{@{}c@{}}Isolate \&\\ Throttle\end{tabular} &
                \begin{tabular}[c]{@{}c@{}}Burst\\ Splitter\end{tabular} &
                \multicolumn{1}{c}{\begin{tabular}[c]{@{}c@{}}Meta\\ Buffer\end{tabular}} &
                \multicolumn{1}{c!{\color{gray}\vrule}}{\begin{tabular}[c]{@{}c@{}}Write\\ Buffer\end{tabular}} &
                \begin{tabular}[c]{@{}c@{}}Tracking\\ counters\end{tabular} &
                \begin{tabular}[c]{@{}c@{}}Region\\ Decoders\end{tabular} \\ \midrule

                Addr Width~\tnote{a}~\tnote{b} &
                \multicolumn{1}{c!{\color{gray}\vrule}}{0} & 
                \multicolumn{1}{c!{\color{gray}\vrule}}{0} &       
                0 &        
                0 &        
                20.6 &  
                3.5 &  
                49.3 & 
                38.1 &
                \multicolumn{1}{c!{\color{gray}\vrule}}{0} &  
                0 &
                20.8 \\

                Data Width~\tnote{a}~\tnote{b} &
                \multicolumn{1}{c!{\color{gray}\vrule}}{0} & 
                \multicolumn{1}{c!{\color{gray}\vrule}}{0} &       
                0 &        
                0 &        
                0 &  
                2.7 &  
                1.5 &
                0 &
                \multicolumn{1}{c!{\color{gray}\vrule}}{0} &  
                0 &
                0 \\

                Num Pending~\tnote{d} &
                \multicolumn{1}{c!{\color{gray}\vrule}}{0} & 
                \multicolumn{1}{c!{\color{gray}\vrule}}{0} &       
                0 &        
                0 &        
                0 &  
                9.0 &  
                729.4 & 
                0 &
                \multicolumn{1}{c!{\color{gray}\vrule}}{0} &  
                0 &
                0 \\

                Buffer Depth~\tnote{d} &
                \multicolumn{1}{c!{\color{gray}\vrule}}{0} & 
                \multicolumn{1}{c!{\color{gray}\vrule}}{0} &       
                0 &        
                0 &        
                0 &  
                0 &  
                0 & 
                0 &
                \multicolumn{1}{c!{\color{gray}\vrule}}{0} &  
                0 &
                0 \\

                Storage Size~\tnote{f,g} &
                \multicolumn{1}{c!{\color{gray}\vrule}}{0} & 
                \multicolumn{1}{c!{\color{gray}\vrule}}{0} &       
                0 &        
                0 &        
                0 &  
                0 &  
                0 & 
                0 &
                \multicolumn{1}{c!{\color{gray}\vrule}}{264.4} &  
                0 &
                0 \\

                Constant~\tnote{h} &
                \multicolumn{1}{c!{\color{gray}\vrule}}{260.6} & 
                \multicolumn{1}{c!{\color{gray}\vrule}}{83.5} &       
                24.6 &        
                1319.6 &        
                0 &  
                267.1 &  
                4835.0 & 
                1309.7 &
                \multicolumn{1}{c!{\color{gray}\vrule}}{11.4} &  
                1928.5 &
                0 \\

                \bottomrule
                                        
            \end{tabular}

            \begin{tablenotes}[para, flushleft]
                \item[a] In [bit]
                \item[b] Evaluated \SIrange{32}{64}{\bit}
                \item[d] Evaluated \SIrange{2}{16}{elements}
                \item[f] Product of \emph{Buffer Depth} and \emph{Data Width}
                \item[g] Evaluated \SIrange{256}{8192}{\bit}
                \item[h] Base area independent of params
            \end{tablenotes}
        \end{threeparttable}
    }
\end{table*}

In this section, we provide both in-system and \gls{ip}-level evaluation of {\axirt}. %
In \Cref{sec:eval:sys}, we integrate the unit in Cheshire, a 64-bit, open-source Linux-capable RISC-V \gls{soc}~\cite{10163410}. We synthesize the \gls{soc} to evaluate the {\axirt} area overhead in the system.
To assess system-level functional performance, we implement the design on an \gls{fpga} and use a representative benchmark from the {MiBench} automotive suite~\cite{990739} and the resources available in the RISC-V \gls{soc}. %
In \Cref{sec:eval:ip}, we perform synthesis on the standalone \gls{rtunit}, traversing its design space and deriving an area model and the timing dependency at varying configurations. %
For gate-level assessment, we use {GlobalFoundries}' \SI{12}{\nano\metre} node with a 13-metal stack and 7.5-track standard cell library in the typical corner. %
We synthesize the designs using {Synopsys} {Design} {Compiler} {NXT} in topological mode to account for place-and-route constraints, congestion, and physical phenomena. %

\subsection{In-system Evaluation}
\label{sec:eval:sys}

\begin{figure}[t]
    \centering
    \includegraphics[width=\columnwidth]{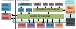}
    \caption{%
        Integration into the {Cheshire} \gls{soc}, \textcolor{ieee-bright-orange-80}{$\blacksquare$} denotes the \glspl{rtunit}. %
        Without limiting generality, we show one \gls{dsa} connected to {Cheshire}.
    }
    \label{fig:cheshire}
\end{figure}

We integrate {\axirt} into {Cheshire}~\cite{10163410}, a 64-bit Linux-capable \gls{soc} designed to host \gls{axi4}-based \glspl{dsa}. %
As shown in \Cref{fig:cheshire}, a \gls{rtunit} is connected to every critical manager used during runtime: the {CVA6} core, the \gls{soc}-level \gls{dma} engine, and the \glspl{dsa} manager ports. %
For the evaluation, we configure {Cheshire} at design time, i.e., before \gls{fpga} or ASIC mapping.
We select one \gls{dsa} port connected to an accelerator's \gls{dma} engine~\cite{benz2023highperformance} and add a dedicated \gls{spm} to serve as local memory to fully utilize the \gls{llc} as a cache. %
The \gls{rtunit} is parameterized with eight \emph{pending transactions}, a write \emph{buffer depth} of sixteen elements, and two subordinate address regions for each manager. 
Only one region is used during the evaluation, encompassing the \gls{llc}.
The \glspl{rtunit} are configured through a dedicated configuration register file connected and mapped into the \gls{soc}'s address space, see \Cref{fig:cheshire}. %
This interface is protected with the \emph{bus guard} introduced in \Cref{sec:cfg}. %
After \gls{fpga} mapping, {CVA6} claims the configuration registers and initializes the \glspl{rtunit} at an early stage of the boot flow, %
i.e., \emph{before} runtime operation.%

We select \emph{Susan} as a function task for {CVA6} from the {MiBench} suite. 
\emph{Susan} features the highest memory intensity and is the most representative {MiBench} automotive benchmark to investigate interference effects in the interconnect bus. %
The application is executed on top of Linux from the DDR3 \gls{dram} available on the \gls{fpga} as main memory and accessed through {Cheshire}'s \gls{llc}. %
As a baseline, the performance of \emph{Susan} on {CVA6} as a \emph{single source} of memory transactions, without the \gls{dsa} \gls{dma} transferring data, highlighted by the \emph{grey dashed line} in \Cref{fig:perf}. %
In this \emph{single source} case, accesses by {CVA6} take at most \emph{eight} cycles to be served by the interconnect, assuming the \gls{llc} is hot. %
During the execution of \emph{Susan} on the core, we program the \gls{dsa} \gls{dma} to perform the worst-case access pattern: \emph{double-buffering} full-length data bursts of \emph{256 beats} between the system's \gls{llc} and the \gls{dsa}'s local \gls{spm}. 
Due to the transaction-granular \gls{rr} arbitration, this leads to the worst-case disturbance and contention, as \emph{every} core access is delayed by \emph{256} cycles. %
Given this worst-case scenario of \emph{uncontrolled} contention (denoted \emph{without reservation} in \Cref{fig:perf-perf}), {CVA6} requires a minimum of \emph{264} cycles for every memory access, resulting in fewer than \SI{0.7}{\%} of the \emph{single-source} performance.

\subsubsection{Influence of Fragmentation Size}
To assess the beneficial impact of {\axirt} under this high contention, we activate the \glspl{rtunit} varying the fragmentation length between \emph{256}, where the unit lets all bursts pass without fragmentation (this corresponds to the \emph{uncontrolled} scenario with high contention \emph{without reservation}), and \emph{one}, where all bursts are decomposed in single-world-granular transfers, restoring fairness from managers launching long bursts. %
As \Cref{fig:perf} shows, we can improve the execution time from \SI{0.7}{\%} to \SI{68.2}{\%} of the \emph{single-source} baseline performance at a fragmentation granularity of \emph{one}. %
We can reduce {CVA6}'s memory access latency from \emph{264} to less than \emph{ten} cycles, only \emph{two} cycles higher than in the \emph{single-source} scenario. %
One cycle of latency is introduced by the \gls{rtunit}, and one can arise from the interference given by the fragmented \gls{dma} transfer. %
In this experiment, we select a very large period and an equal budget for both {CVA6} and the \gls{dma} to focus on the impact of fragmentation size on traffic fairness. %

\subsubsection{Influence of Budget Distribution}
\begin{figure}[t]
    \centering
    \subfloat[\label{fig:perf-perf}]{{\includegraphics[height=3.2cm]{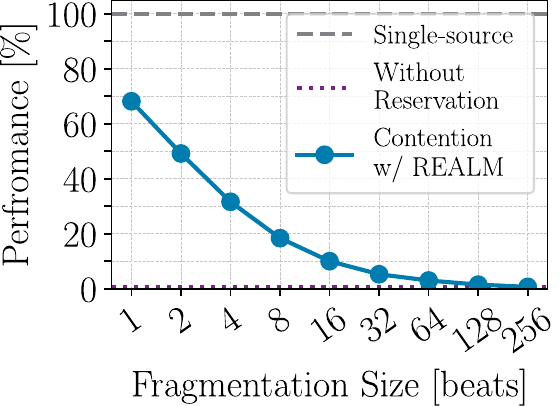}}}%
    \quad
    \subfloat[\label{fig:perf-peri}]{{\includegraphics[height=3.2cm]{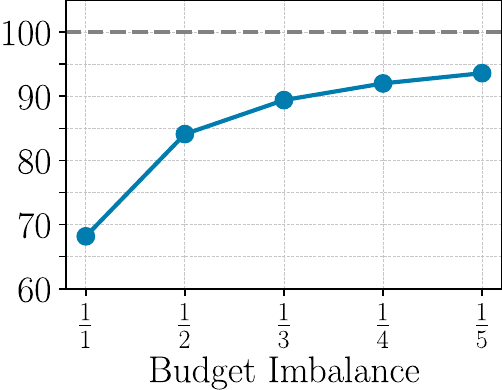}}}%
    \caption{%
        (a) Performance of \emph{Susan} on {CVA6} under \gls{dsa} \gls{dma} contention at varying transfer fragmentation in beats. A fragmentation size of \emph{one} provides the most fair scenario and close to ideal performance. %
        \emph{Single-source} denotes the performance without interference. %
        \emph{Without reservation} gives the performance in the case of \textbf{uncontrolled} contention. %
        (b) Performance achieved by varying the budget imbalance between the core and the \gls{dma}. The fragmentation size is set to one, maximizing fairness under large contention. %
        Traversing the x-axis from left to right, less budget is reserved for the \gls{dma}, closing the gap in the core performance under contention compared to the \emph{single-source} scenario.
    }
    \label{fig:perf}
\end{figure}

Moreover, we observe that near-ideal (\emph{dashed line}) core performance can be achieved by reducing the balance between the \gls{dma} and the core budget, as depicted in \Cref{fig:perf-peri}. %
Decreasing the \gls{dma} budget reduces the likelihood of \gls{dma} contention, further decreasing {CVA6}'s memory access latency. %
We configure the fragmentation size to \emph{one} beat, i.e., the optimal fairness condition shown in \cref{fig:perf-perf}, with an equal budget for both managers and select a short period of \emph{1000} clock cycles. %
The \gls{dma}'s budget is then reduced from \SI{8}{\kilo\byte} (denoted by \emph{$\sfrac{1}{1}$}) to \SI{1.6}{\kilo\byte} (\emph{$\sfrac{1}{5}$}) in equal steps. %

\subsubsection{In-system {\axirt} Area Overhead}
To assess {\axirt}'s resource overhead at the system level, we synthesize {Cheshire} in the same configuration shown in \Cref{fig:cheshire}. %
Overall, the \glspl{rtunit} increase the \gls{soc}'s area by \SI{83.6}{\kGE} or \SI{2.45}{\%}, as shown in \Cref{tab:area-soc}.
Introducing {\axirt} did not reduce {Cheshire}'s maximum operating frequency. %

\subsection{IP-level Evaluation}
\label{sec:eval:ip}

To facilitate area and timing estimates when integrating into different systems, promoting fair comparison with other works, we provide a detailed area model of {\axirt} in \Cref{tab:ooc-area}. %
The data is grouped into two categories: \emph{configuration register file} and the \emph{\gls{rtunit}}, grouped into sub-categories \emph{per system}, \emph{per unit}, and \emph{per unit and region}. %
To estimate the area of an {\axirt} system, the individual unit's area contributions are multiplied by the parameter value and summed up. %
The area estimates in \Cref{tab:ooc-area} are provided at \SI{1}{\giga\hertz}, the target frequency of {CVA6} in this node. %
{\axirt} can easily achieve clock speeds exceeding \SI{1.5}{\giga\hertz} without adding additional cuts. %
If an application demands higher frequencies, pipeline stages can be introduced into the \gls{rtunit}. %

\section{Conclusion}
\label{sec:concl}
This paper presents {\axirt}, a lightweight, technology-independent interconnect bus extension for heterogeneous real-time \glspl{soc} based on \gls{axi4}, the de facto standard for non-coherent, on-chip interconnects. %
We implement {\axirt} in a \gls{sota} Linux-capable RISC-V \gls{soc}, only introducing \SI{2.45}{\%} of area overhead at isofrequency with the original design. %
{\axirt} is beneficial to restore bandwidth fairness and limits bandwidth utilization in systems with independent actors issuing high-congestion transactions on a single crossbar, such as \glspl{ha} in heterogeneous systems. %
In our experiments with two critical managers contending the system bus (a processor running a representative benchmark from the {MiBench} suite and a \gls{dsa}'s \gls{dma} engine), {\axirt} can restore the core performance to \SI{68.2}{\%} of the isolated scenario with equal budget distribution across the managers, and near-ideal performance (above \SI{95}{\%}) when distributing the budget in favor of the core, improving the worst-case memory access latency from \emph{264} to below \emph{eight} cycles.
We provide an area model of {\axirt} in a \SI{12}{\nano\metre} node to facilitate rapid exploration and allow fair comparisons. %
Thanks to {\axirt}'s modularity, use cases beyond real-time embedded computing could be targeted: {\axirt} %
could be used in multi-tenant smart {NICs}~\cite{osmosis} to enforce guarantees on shared resource usages. %
We provide {\axirt}'s \gls{rtl} free and open-source. %

\section*{Acknowledgments}
\label{sec:ack}
\ifx\blind\undefined
    We thank %
    Florian Zaruba, %
    Andreas Kurth, %
    Wolfgang Rönninger, %
    Michael Rogenmoser, %
    Paul Scheffler,
    and %
    Sergio Mazzola, %
    for their valuable contributions to the research project. %
    This work was supported in part through the TRISTAN project that has received funding from the Key Digital Technologies Joint Undertaking (KDT JU) under grant agreement nr. 101095947. %
    The KDT JU receives support from the European Union's Horizon Europe's research and innovation programme and Austria, Belgium, Germany, Finland, France, Israel, Italy, Netherlands, Poland, Romania, Switzerland and Turkey. %
\else
    \textit{Acknowledgments omitted for blind review.}
\fi

%

%


\end{document}